\documentclass[a4paper]{article}
\usepackage[14pt]{extsizes}
\usepackage[english]{babel}
\usepackage[intlimits]{amsmath} 
\usepackage{hyperref}
\usepackage{indentfirst}
\usepackage{geometry} 
\usepackage{xfrac}
\usepackage{setspace}
\providecommand{\keywords}[1]{\textbf{Keywords: } #1}

\begin{document}
\title{A new way to accelerate the
D-MORPH method to search for optimal quantum control}
\author{Konstantin Zhdanov\thanks{Department of applied mathematics and control processes, Saint-Petersburg state university, Russia}}
\date{}

\maketitle
\begin{abstract}
The paper introduces new corrections of different orders of smallness to the ~\\D-MORPH method by using the full form of the derivative of the exponential map, defined on a Lie algebra, to search for the optimal control of a quantum system that implements a desired unitary evolution. The inclusion of such corrections, which take into account information about different system Hamiltonian’s commutators, results in faster optimal control’s finding, even compared to the improved version of the method published earlier by the author.
\end{abstract}

\keywords{quantum systems, optimal control, evolution operator, \newline Lie algebras, quantum computations}

\section*{Introduction}
The D-MORPH method~\cite{Moore1, Moore2, Moore-Tibetts, Riviello1, Riviello2} is one of the most frequently used methods for constructing an optimal control of quantum systems that minimizes a given objective function. The D-MORPH method is based on numerical integration of a specially constructed system of ordinary differential equations with independent variable \(s\), which represents the optimization's progress towards the minimum of the objective function, using the variable-step Runge--Kutta's method of order four. That process usually takes a long time to reach a small vicinity of the minimum. As was shown author's previous paper~\cite{Zhdanov}, the D-MORPH method can be speeded up with use of information about different quantum system Hamiltonian's commutators.

In this paper a new method for obtaining corrections of different orders of smallness to the D-MORPH method is presented, such that they allow a speed-up of quantum optimal control search by using the complete form of the derivative of the exponential map defined on a Lie algebra. The corrections obtained resemble those found by the author earlier in~\cite{Zhdanov}, hence these two forms of corrections are compared with each other by numerically implementing a quantum gate, and it's shown that the use of the new form results in shorter times spent to solve the underlying system.
 
\section*{The previously obtained method}
To implement a desired unitary operator \(U_D\) at a moment \(T\) in a \(N\)-level quantum system with the Hamiltonian \(H(t)=H_0 + \sum_{k=1}^M \epsilon_k(t) H_k\) subject to \(M\) controls \(\{\epsilon_k\}_{k=1}^M\) the D-MORPH method requires solving of the system
\begin{equation}\label{DMORPH_orig}
\frac{\mathrm{d}\epsilon_k^l (s)}{\mathrm{d} s} = - \frac{\partial J}{\partial \epsilon_k^l (s)},\quad l=1,\ldots,L,\quad k=1,\ldots,M,
\end{equation}
on the interval \([0,S]\) divided into \(L\) equal-length subintervals \([t_{l-1}, t_l]\), where \(J=0.5-\operatorname*{Re} \operatorname*{Tr}\left( U^*_D U(T, 0)\right)/(2N)\) --- an objective function to be minimized, \(\epsilon_k^l\) --- constant controls' values on \([t_{l-1}, t_l]\), \(s\) --- the parameter representing the progress towards the minimum, \(\operatorname*{Tr}(\ldots)\) --- the trace operation, \(U^*_D\) --- the Hermitian adjoint of \(U_D\), \(U(T,0)\) --- the system's evolution operator on \([0, T]\).

In the previous paper~\cite{Zhdanov} the following corrections of different orders to D-MORPH were obtained by integrating system~\eqref{DMORPH_orig} with respect to time \(t\):
\begin{equation}\label{Corrections_old_full}
\frac{\mathrm{d}\epsilon_k^l}{\mathrm{d} s} = \frac{1}{2N}\operatorname*{Im} \operatorname*{Tr}\left(U_D^* U(T, t_{l-1}) \left[\sum_{n=0}^{\infty} \frac{(\imath \Delta t)^n}{(n+1)!} \operatorname{ad}^n_H \circ H_k \right] U(t_{l-1}, 0)\right),
\end{equation}
where \(H = H_0 + \sum_{k=1}^M \epsilon_l^k H_k\) --- the constant Hamiltonian on \([t_{l-1}, t_l]\), \(\operatorname{ad}_H\circ H_k = [H, H_k] = H H_k - H_k H\) --- the commutator of matrices \(H\) and \(H_k\), \(\operatorname{ad}^n_H = \operatorname{ad}_H \operatorname{ad}^{n-1}_H\). The original D-MORPH can be obtained by letting \(\Delta t = 0\), i.~e. by taking the first term only in the series~\eqref{Corrections_old_full}. As it was shown in~\cite{Zhdanov}, method~\eqref{Corrections_old_full} takes less time to reach the minimum of the objective function and requires a smaller integration interval \([0, S]\) compared to D-MORPH.
\section*{The new way to obtain corrections}
There is one more way to obtain analogous corrections to D-MORPH based on more accurate and theoretically correct calculation of the derivative of the objective function \(\sfrac{\partial J}{\partial \epsilon_k^l}\). Similar to the paper~\cite{deFouquieresa},  the full form of the derivative of the exponential matrix \(\exp(-\imath \Delta t H)\) with respect to \(\epsilon_k^l\) is used, where the exponential matrix is the evolution operator of a quantum system with the Hamiltonian \(H\) on an interval of length \(\Delta t\). This form of the derivative has been known for a long time in the general theory of Lie groups and algebras~\cite[p.~15]{Rossmann} and in the case of a quantum system can be written as
\begin{multline*}
\frac{\partial}{\partial \epsilon_k^l} \exp(-\imath \Delta t H) = \exp(-\imath \Delta t H) \frac{1-\exp(-\operatorname{ad}_{-\imath \Delta t H})}{\operatorname{ad}_{-\imath \Delta t H}}\circ \left(-\imath \Delta t H_k\right) =\\
= -\imath \Delta t \exp(-\imath \Delta t H) \sum_{n=0}^\infty \frac{(\imath \Delta t)^n}{(n+1)!} \operatorname{ad}^n_H\circ H_k.
\end{multline*}

In the author's previous paper~\cite{Zhdanov}, only the term with \(n=0\) was used, which was similar to using the classic formula of the derivative of the scalar exponential function. It can be easily seen that the error of such approximation was \(O(\Delta t)\) and it could have affected the accuracy of numerical computations, especially in the case of large steps \(\Delta t\). Therefore, the full form of the derivative is employed in this paper to potentially make the optimization process quicker and more accurate. After substituting this form into system~\eqref{DMORPH_orig}, a new system of differential equations is obtained as
\begin{equation}\label{Corrections_new_full}
\frac{\mathrm{d}\epsilon_k^l}{\mathrm{d} s} = \frac{\Delta t}{2N}\operatorname*{Im} \operatorname*{Tr}\left(U_D^* U(T, t_{l-1}) \left\{\sum_{n=0}^{\infty} \frac{(\imath \Delta t)^n}{(n+1)!} \operatorname{ad}^n_H \circ H_k \right\} U(t_{l-1}, 0)\right).
\end{equation}
This system looks similar to~\eqref{Corrections_old_full} obtained earlier in~\cite{Zhdanov}, except for the presence of factor \(\Delta t\) and for the fact that the new system more accurately describes the derivative. It's expected that method~\eqref{Corrections_new_full} will give more precise results than method~\eqref{Corrections_old_full}, which in turn gives more accurate results than D-MORPH. Furthermore, it's expected that the new method~\eqref{Corrections_new_full} will find near-optimal controls faster than method~\eqref{Corrections_old_full}.
\section*{Numerical experiment}
In order to confirm an increase in accuracy and speed of method~\eqref{Corrections_new_full} compared to method~\eqref{Corrections_old_full}, only the first two terms of the series were employed (\(n=0\) and \(n=1\)) for solving a problem in the area of quantum computations --- implementing the controlled NOT~\cite[p.~21]{Nielsen} quantum gate (CNOT) in a quantum system consisting of two spin-\(\sfrac{1}{2}\) particles (\(N=4\)), which is described by the dimensionless Hamiltonian
\[
H = \sum_{i=1}^2 S_z^i \omega_i + \sum_{k=1}^2 \epsilon_k^l S_x^k + C_x^{(12)} S_x^1 S_x^2 + C_y^{(12)} S_y^1 S_y^2 + C_z^{(12)} S_z^1 S_z^2,
\] 
where \(\omega_1=20\), \(\omega_2=30\), \(C_x^{(12)}=110\), \(C_y^{(12)}=120\), \(C_z^{(12)}=130\), \(S_i^2 = I\otimes S_i\), \(S_i^1 = S_i \otimes I\), 
\[
S_x = \begin{pmatrix}
0 & 1\\
1 & 0
\end{pmatrix},
S_y = \begin{pmatrix}
0 & -\imath \\
\imath & 0
\end{pmatrix},
S_z = \begin{pmatrix}
1 & 0\\
0 & -1
\end{pmatrix},
U_D = \mathrm{e}^{\frac{\imath \pi}{4}}\begin{pmatrix}
1 & 0 & 0 & 0\\
0 & 1 & 0 & 0\\
0 & 0 & 0 & 1\\
0 & 0 & 1 & 0
\end{pmatrix}.
\]

Therefore the following two models were compared:
\begin{equation} \label{Corrections_old_1st}
\frac{\mathrm{d}\epsilon_k^l}{\mathrm{d} s} = \frac{1}{2N}\operatorname*{Im} \operatorname*{Tr}\left(U_D^* U(T, t_{l-1}) \left\{H_k + \frac{\imath \Delta t}{2} \left[H,H_k\right] \right\} U(t_{l-1}, 0)\right),
\end{equation}
\begin{equation} \label{Corrections_new_1st}
\frac{\mathrm{d}\epsilon_k^l}{\mathrm{d} s} = \frac{\Delta t}{2N}\operatorname*{Im} \operatorname*{Tr}\left(U_D^* U(T, t_{l-1})  \left\{H_k + \frac{\imath \Delta t}{2} \left[H,H_k\right] \right\} U(t_{l-1}, 0)\right).
\end{equation}
 
These systems were numerically solved with MATLAB's ode45 method~\cite{Shampine} on a quad-core processor Intel Core i7 2.20 GHz with 12 Gb RAM. The initial control fields were taken equal to zero everywhere on \([0,T]\) and the system's parameters were taken as follows: \(T=5, 10\); \(L=150, 300\); \(S=5000\). The numerical integration was stopped when either \(S=5000\) was reached or the objective function \(J < 10^{-7}\), i.~e. when the error in the CNOT gate's implementation became less than \(10^{-7}\). The comparison results are summarized in the following table.
\begin{center}
\begin{tabular}{ *{9}{|c} |}
\hline
& \multicolumn{2}{|c|}{\(T=10\)} & \multicolumn{2}{|c|}{\(T=10\)} & \multicolumn{2}{|c|}{\(T=5\)} & \multicolumn{2}{|c|}{\(T=5\)}\\
& \multicolumn{2}{|c|}{\(L=300\)} & \multicolumn{2}{|c|}{\(L=150\)} & \multicolumn{2}{|c|}{\(L=300\)} & \multicolumn{2}{|c|}{\(L=150\)}\\
\hline
Method & \eqref{Corrections_old_1st} & \eqref{Corrections_new_1st} & \eqref{Corrections_old_1st} & \eqref{Corrections_new_1st} & \eqref{Corrections_old_1st} & \eqref{Corrections_new_1st} & \eqref{Corrections_old_1st} & \eqref{Corrections_new_1st}\\
\hline
Final S & 70 & 2089 & 73 & 1097 & 81.1 & 4866.5 & 117.8 & 3532.2\\
\hline
Time, sec. & 348 & 317 & 174.8 & 158.8 & 240.8 & 225.7 & 99 & 94.3\\
\hline
Max step & 0.19	& 5.6 & 0.14 & 2.1 & 0.3 & 19.4 & 0.4 & 13\\
\hline
\end{tabular}
\end{center}

It can be clearly seen from the table that in comparison with method~\eqref{Corrections_old_1st}, method~\eqref{Corrections_new_1st} proposed in this paper always reached the desired precision more quickly but using a larger interval described in the table by the final S value where the numerical integration was stopped. It's worth mentioning that methods~\eqref{Corrections_old_full} and~\eqref{Corrections_new_full} with only one term (\(n=0\)) were compared to each other as well, and the new formula performed much worse than the original \newline D-MORPH method --- it required larger computation times and larger intervals to reach the desired precision than D-MORPH, and worst of all, sometimes this method didn't converge at all.

\section*{Conclusion}
A new way to obtain corrections of different orders of smallness to the D-MORPH method for numerical construction of optimal quantum control was presented in the paper. The corrections obtained turned out to be analogous to the ones derived earlier by the author~\cite{Zhdanov}, except for the presence of a factor proportional to the time step used to divide the interval of the system's evolution and for the fact that the new method more accurately describes the derivative of the objective function. By solving the problem of the CNOT gate's implementation in a quantum system composed of two spin-\(\sfrac{1}{2}\) particles, it was shown that the new method could actually speed up the optimization process compared to both the original D-MORPH and the previously obtained method~\cite{Zhdanov} --- the inclusion of only one additional correction term allowed the new method to implement the desired logical quantum operation more quickly. It's noteworthy that in the case of comparison between the original method and the new method without any correction terms (\(n=0\)), the latter almost always performed much worse and sometimes didn't converge to the optimum at all --- information about the system Hamiltonian's commutators turns out to be crucial for the method to work. If no such information is available, then the best option is to use the original D-MORPH method. Thus the use of the theoretically correct form of the derivative of the exponential map to derive the corrections allows further improvement to accuracy and performance of the method for finding optimal quantum control, which should be beneficial when using computers with moderate performance.

\bibliographystyle{plain}

\begin{thebibliography}{9}
\bibitem{Zhdanov} Zhdanov K.~E. An improvement of D-MORPH method for finding quantum optimal control / K.~E. Zhdanov // International research journal. 2016. Vol. 6(48). P. 94--99. DOI: 10.18454/IRJ.2016.48.057 [in Russian]
\bibitem{deFouquieresa} de Fouquieresa P. Second order gradient ascent pulse engineering / P. de Fouquieresa, S.~G. Schirmera, S.~J. Glaserb, et. al. // Journal of Magnetic Resonance. 2011. Vol. 212(2). P. 412--417.
\bibitem{Moore1} Moore K.~W. Search complexity and resource scaling for the quantum optimal control of unitary transformations /  K.~W. Moore, R. Chakrabarti, G. Riviello, et. al. // Phys. Rev. A. 2011. Vol. 83(1).
\bibitem{Moore2} Moore K.~W. Exploring constrained quantum control landscapes / 
K.~W. Moore, H. Rabitz // The Journal of Chemical Physics. 2012. Vol. 137(13).
\bibitem{Moore-Tibetts} Moore Tibbetts K. Exploring the trade-off between fidelity and time optimal control of quantum unitary transformations / K. Moore Tibbetts, C. Brif, M.~D. Grace, et. al. // Phys. Rev. A. 2012. Vol. 86(6). 
\bibitem{Nielsen} Nielsen M.~A. Quantum Computation and Quantum Information: 10th Anniversary Edition / M.~A. Nielsen, I.~L. Chuang. New York: Cambridge University Press, 2010. 702 p.
\bibitem{Riviello1} Riviello G. Searching for quantum optimal controls in the presence of singular critical points / G. Riviello, C. Brif, R. Long, et. al. // Phys. Rev. A. 2014. Vol. 90(1).
\bibitem{Riviello2} Riviello G. Searching for quantum optimal controls under sever constraints / G. Riviello, K. Moore Tibbetts, C. Brif, et. al. // Phys. Rev. A. 2015. Vol. 91(4).
\bibitem{Rossmann} Rossmann W. Lie Groups: An Introduction Through Linear Groups / 
W. Rossmann. New York: Oxford University Press, 2006. 265 p.
\bibitem{Shampine} Shampine L.~F. The MATLAB ODE Suite / L.~F. Shampine, M. W. Reichelt // SIAM Journal on Scientific Computing. 1997. Vol. 18(1). \newline P. 1--22.

\end{thebibliography}

\end{document}